\documentclass{PoS}

\title{Applications of the loop--tree duality}

\ShortTitle{Applications of the loop--tree duality}

\author{\speaker{Germ\'an Rodrigo}, F\'elix Driencourt-Mangin, Germ\'an F. R. Sborlini \\ 
        Instituto de F\'isica Corpuscular, Universitat de Val\`encia -- Consejo Superior de Investigaciones Cient\'ificas,
        Parc Cient\'ific, E-46980 Paterna, Valencia, Spain \\
        E-mail: \email{german.rodrigo@csic.es, felix.dm@ific.uv.es, german.sborlini@ific.uv.es}}

\author{Roger Jos\'e Hern\'andez-Pinto\\
        Facultad de Ciencias F\'isico-Matem\'aticas, Universidad Aut\'onoma de Sinaloa, Ciudad Universitaria, 
        CP 80000, Culiac\'an, Sinaloa, M\'exico\\
        E-mail: \email{roger@uas.edu.mx}}

\abstract{We describe a new method to perform NLO calculations, combining real and virtual amplitudes at the
integrand level, with a fully local compensation between them in the IR, and between the virtual integrand and 
properly defined counter-terms in the UV, in such a way that physical observables can be computed  in 4 dimensions. 
One of the advantages of the method is that all the scattering amplitudes are integrated simultaneously, without the need 
for tensor reduction, or projection onto sets of master integrals. As such, it could offer great progress in the 
automation of NLO calculations, where the actual bottle-necks are the complexity of
the analytical calculations for multi-leg processes, and the numerical stability of the result.}

\FullConference{Loops and Legs in Quantum Field Theory\\
		24-29 April 2016\\
		Leipzig, Germany}

\def\beq{\begin{equation}}
\def\eeq{\end{equation}}
\def\bea{\begin{eqnarray}}
\def\eea{\end{eqnarray}}
\def\beqn{\begin{eqnarray}} \def\eeqn{\end{eqnarray}}
\def\beeq{\begin{eqnarray}}
\def\eeeq{\end{eqnarray}}

\def\ep{\epsilon}

\def\nn{\nonumber}
\def\Eq#1{Eq.~(\ref{#1})}

\def\td#1{\tilde{\delta}\left(#1\right)}

\newcommand{\la}{\langle}
\newcommand{\ra}{\rangle}

\def\M#1{{\cal M}^{(#1)}} 
\def\MD#1{{\cal M}^{(#1) \, \dagger}}

\newcommand\as{\alpha_{\mathrm{S}}}

\def\uv{{\rm UV}}

\def\as{\alpha_{\rm S}}

\def\v{{\rm V}}
\def\r{{\rm R}}

\def\xiu{\xi_{1,0}}
\def\xid{\xi_{2,0}}
\def\xit{\xi_{3,0}}

\def\nlo{{\rm NLO}}

\begin{document}

\section{The loop--tree duality}

The standard approach to perform perturbative calculation in QCD relies in the application of the subtraction formalism. 
There are several variants of the subtraction method at NLO and beyond~\cite{Kunszt:1992tn,Frixione:1995ms, Catani:1996jh, Catani:1996vz, GehrmannDeRidder:2005cm,Catani:2007vq,Czakon:2010td,Bolzoni:2010bt,DelDuca:2015zqa,Boughezal:2015dva,Gaunt:2015pea}, 
which involve treating separately real and virtual contributions.
Besides the computational difficulty related to the fact that the final-state phase-space of the different contributions 
involves different number of particles,  building these counter-terms represents a challenge and introduces a potential 
bottleneck to efficiently carry out the infrared (IR) subtraction for multi-leg multi-loop processes.

With the aim of eluding the introduction of IR counter-terms, we have developed an alternative idea based on the application of the loop--tree duality (LTD) \cite{Catani:2008xa,Bierenbaum:2010cy,Bierenbaum:2012th,Bierenbaum:2013nja,Buchta:2014dfa,Buchta:2015xda,Buchta:2015wna,Hernandez-Pinto:2015ysa,Sborlini:2015uia,Sborlini:2016fcj,Sborlini:2016gbr}~\footnote{See also Ref.~\cite{Seth:2016hmv}.}.
The LTD theorem establishes that loop scattering amplitudes, at one-loop and beyond, 
can be expressed as a sum of phase-space integrals (i.e. the so-called \textit{dual integrals}) with additional physical particles. 
Dual integrals and real-radiation contributions exhibit a similar structure, and can be combined at integrand level. 
For example, the dual representation of the scalar one-loop integral over $N$ Feynman propagators, $G_F(q_i)$,
is the sum of $N$ dual integrals~\cite{Catani:2008xa,Bierenbaum:2010cy}:
\bea
\label{oneloopduality}
L^{(1)}( \{p_i\}) 
&=& \int_\ell \, \prod_{i\in\alpha_1} G_F(q_i) = - \sum_{i\in \alpha_1} \, \int_{\ell} \; \td{q_i} \,
\prod_{j \in \alpha_1, j\neq i} \,G_D(q_i;q_j)~, \quad \alpha_1=\{1,\cdots, N\}~,
\eea 
where
\beq
G_D(q_i;q_j) = \frac{1}{q_j^2 -m_j^2 - \imath \, 0 \, \eta \, k_{ji}}
\eeq
are the so-called dual propagators, with $\eta$ a {\em future-like} vector, $\eta^2 \ge 0$, 
with positive definite energy $\eta_0 > 0$.
The delta function $\td{q_i} \equiv 2 \pi \, \imath \, \theta(q_{i,0}) \, \delta(q_i^2-m_i^2)$
sets the internal lines on-shell by selecting the pole of the Feynman propagators
with positive energy $q_{i,0}$ and negative imaginary part. The sign of the $\imath\, 0$ prescription 
depends now on $k_{ji} = q_j-q_i$, which indeed at one-loop
is a function of the external momenta $p_i$ only. 
Having different $\imath\, 0$ prescriptions for each cut, far from introducing an extra difficulty, 
is a necessary condition for the consistency of the method. 
Actually, it is needed to demonstrate the main property of LTD, 
namely the partial cancellation of singularities among different dual integrands 
and the causal interpretation of the remaining ones~\cite{Buchta:2014dfa}.

Regular loop integrals become singular at the on-shell hyperboloids (light-cones for massless propagators) of the 
Feynman propagators, $G_F^{-1}(q_i)=0$. The first advantage of LTD is that integration over the 
loop momentum is directly set on top of the forward on-shell hyperboloids (positive energy modes). 
Singularities in the loop three-momentum space appear only in the intersection with other forward or 
backward (negative energy modes) on-shell hyperboloids. However, singularities in the intersection 
of forward on-shell hyperboloids cancel among different dual contributions; the change of sign of the modified 
$\imath \,0$ prescription is crucial to enable this behaviour. Only singularities in the intersection of forward with 
backward on-shell hyperboloids (FB) remain, and even more important, they are constrained to a compact region 
of the loop-three momentum space and are easily reinterpreted in terms of causality. 
From a physical point of view, FB singularities take place when the on-shell virtual particle interacts with another 
on-shell virtual particle after the emission of outgoing on-shell radiation. 
The direction of the internal momentum flow establishes a natural causal ordering, 
and this interpretation is consistent with the Cutkosky rule. 

The compactness of the singular region is crucial to advance one steep further in the 
implementation of the method. Threshold singularities are integrable, and can be integrated 
numerically by contour deformation~\cite{Buchta:2014dfa,Buchta:2015xda,Buchta:2015wna}. 
The treatment of IR singularities is discussed in the next Section.

\section{Four dimensional unsubtraction}
\label{sec:global}

The summation over degenerate soft and collinear states from virtual and real corrections 
is implemented in LTD by a suitable mapping of momenta between the virtual and real kinematics. 
This is possible because the IR singularities are restricted to a compact region of the loop three-momentum,
and makes unnecessary the introduction of IR subtractions for soft and final-state collinear singularities. 
As usual, the NLO cross-section is constructed from the one-loop virtual correction with $m$ partons in the final 
state and the exclusive real cross-section with $m+1$ partons in the final state
\beq
\sigma^\nlo = \int_{m} d\sigma_{\v}^{(1,\r)}+ \int_{m+1} d\sigma_{\r}^{(1)}~,
\eeq
where the virtual contribution is obtained from its dual representation 
\beq
d\sigma_{\v}^{(1,\r)} = \sum_{i=1}^N \int_\ell \,  2 \, {\rm Re} \, \la \M{0}_N|\M{1,\r}_N(\td{q_i}) \ra \, 
{\cal O}_N (\{p_i\}) ~.
\label{eq:nlov}
\eeq
In \Eq{eq:nlov}, $\M{0}_N$ is the $N$-leg scattering amplitude at LO, and $\M{1,\r}_N$ is the renormalised 
one-loop scattering amplitude, which also contains the self-energy corrections of the external legs, even 
if they are massless and then ignored in the usual calculations because their integrated form vanishes. 
The delta function $\td{q_i}$ symbolises the dual contribution with the internal momentum 
$q_i$ set on-shell. The integral is weighted with the function ${\cal O}_N$ that defines a given observable, 
for example the jet cross-section in the $k_T$-algorithm. 
The renormalised amplitude includes appropriate counter-terms that subtract the UV singularities 
locally, as discussed in Ref.~\cite{Sborlini:2016gbr}, 
including UV singularities of degree higher than logarithmic that integrate to zero.
In~\Eq{eq:nlov}, we have also assumed a definite ordering of the external particles that leads to a definite
set of internal momenta. 

The real cross-section is given by 
\bea
\int_{m+1}\, d\sigma^{(1)}_{\r} &=& \sum_{i=1}^N \, \int_{m+1} \, |\M{0}_{N+1}(q_i,p_i)|^2 \, {\cal R}_i(q_i,p_i)\,
{\cal O}_{N+1}(\{p_j'\})~, 
\label{eq:nlor}
\eea 
where the external momenta $p_j'$, the 
phase-space and the tree-level scattering amplitude $\M{0}_{N+1}$ are rewritten in 
terms of the loop three-momentum (equivalently, the internal on-shell loop momenta) and the external 
momenta $p_i$ of the Born process. 

\begin{figure}[thb]
\begin{center}
\includegraphics[width=13cm]{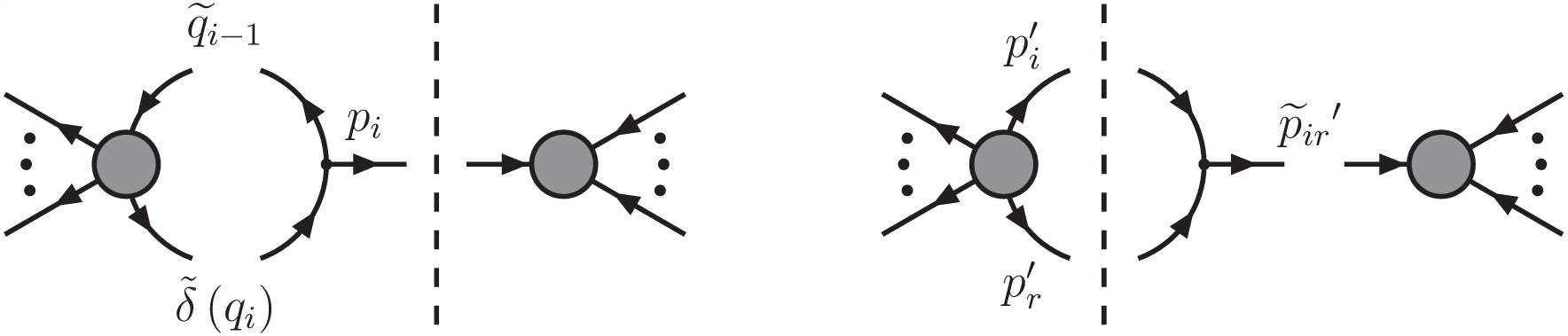}
\caption{\label{fig:collinear}
Interference of the Born process with the one-loop scattering amplitude with 
internal momentum $q_i$ on-shell, $\M{1}_{N}(\td{q_i})\otimes \MD{0}_{N}$ (left), 
and interference of real processes with parton splitting $p_{ir}'\to p_i'+p_r'$:
$\M{0}_{N+1}\otimes \MD{0}_{N+1}(p_{ir}')$ (right). The dashed line represents momentum conservation.
In the soft/collinear limits the momenta $q_{i-1}=q_i-p_i$ and $p_{ir}'$ become on-shell
and the scattering amplitudes factorise.}
\end{center}
\end{figure}

Analogously to the dipole method~\cite{Catani:1996jh,Catani:1996vz}, 
we single out two partons for each contribution in order to construct the momentum 
mapping between the $m$ and $m+1$ kinematics. 
The first parton is the {\it emitter} and the second parton is the {\it spectator}.
Then, the loop three-momentum and the four-momenta of the emitter and the spectator
are used to reconstruct the kinematics of the corresponding real emission cross-section 
in the region of the real phase-space where the twin of the emitter decays to 
two partons in a soft or collinear configuration.  Explicitly, if the momentum of the final-state emitter 
is $p_i$, the internal on-shell momentum prior to the emitter is $q_i$, and $p_j$ is the momentum 
of the final-state spectator, then, the momentum mapping is given by 
\bea
&& p_r'^\mu = q_i^\mu~, \nn \\
&& p_i'^\mu = p_i^\mu - q_i^\mu + \alpha_i \, p_j^\mu~, \qquad \alpha_i = \frac{(q_i-p_i)^2}{2 p_j\cdot(q_i-p_i)}~, \nn \\
&& p_j'^\mu = (1-\alpha_i) \, p_j^\mu~, \nn \\
&& p_k'^\mu = p_k^\mu~, \qquad \qquad \qquad \qquad k \ne i,j~.
\label{eq:multilegmapping}
\eea
The incoming initial-state momenta, $p_a$ and $p_b$, are left unchanged, 
and $p_r'$ is the momentum of the extra radiated particle.  Momentum conservation is 
automatically fulfilled by~\Eq{eq:multilegmapping} because $p_i+ p_j +\sum_{k\ne i,j} p_k = p_i'+p_r'+p_j'+\sum_{k\ne i,j} p_k'$, 
and all the primed final-state momenta are massless and on-shell if the virtual unprimed momenta are also massless. 
The momentum mapping in~\Eq{eq:multilegmapping} is motivated by the general factorisation properties 
in QCD~\cite{Buchta:2014dfa,Catani:2011st}, and it is graphically explained in Fig.~\ref{fig:collinear}.

The emitter $p_i$ has the same flavour as $p_{ir}'$, the twin emitter or parent parton (called emitter in the dipole formalism) 
of the real splitting configuration that is mapped. The spectator $p_j$ is used to balance 
momentum conservation, and has the same flavour in the virtual and real contributions. 
Similar mappings, involving also initial-state momenta, have been presented recently in Ref.~\cite{Seth:2016hmv}.
As for the dipoles, there are alternatives to treat the recoiling momentum, but the option with a single spectator is the simplest one. 

The momentum mapping in \Eq{eq:multilegmapping} is suitable in the region of the loop-momentum space 
where $q_i$ is soft or collinear with $p_i$, and therefore in the region of the real phase-space where 
$p_i'$ and $p_r'$ are produced collinear or one of them is soft. 
In~\Eq{eq:nlov}, we have introduced a complete partition of the real phase-space
\beq
\sum {\cal R}_i (q_i,p_i) = \sum \prod_{jk\ne ir} \theta(y_{jk}' -y_{ir}')   = 1~,
\eeq 
which is equivalent to split the phase-space as a function of the minimal dimensionless two-body invariant $y_{ir}'=s_{ir}'/s$.
Since the real and virtual kinematics are related, the real phase-space partition defines 
equivalent regions in the loop three-momentum space. Notice, however, that we 
have not imposed these constrains in the definition of the virtual cross-section in~\Eq{eq:nlov}.
The actual implementation of the NLO cross-section in a Monte Carlo event generator is a single unconstrained 
integral in the loop three-momentum, and the phase-space with $m$ final-state particles. By virtue of 
the momentum mapping, real corrections are contributing in the region of the loop three-momentum where 
they map the corresponding soft and collinear divergences. This region is compact, 
and it is of the size of the representative hard scale of the scattering process. 
At large loop three-momentum, only the virtual corrections contribute, and their UV 
singularities are subtracted locally by suitable counter-terms. The IR singularities 
are unsubtracted because their cancellation is achieved at the integrand level by combining 
simultaneously virtual and real corrections. The momentum mapping respects the soft and collinear 
limits, therefore the IR cancellation is fulfilled for infrared safe observables and not only for the total cross-section. 
The full calculation can be implemented in four-dimensions, more precisely with the 
Dimensional Regularization (DREG) parameter $\ep=0$. 
 
At NNLO, the total cross-section consists of three contributions 
\beq
\sigma^{\rm NNLO} = \int_{m} d\sigma_{\v\v}^{(2)} + \int_{m+1} d\sigma_{\v\r}^{(2)} + \int_{m+2} d\sigma_{\r\r}^{(2)}~,
\eeq
where the double virtual cross-section $d\sigma_{\v\v}^{(2)}$ receives contributions from 
the interference of the two-loop with the Born scattering amplitudes, and the square 
of the one-loop scattering amplitude with $m$ final-state particles, the virtual-real cross-section 
$d\sigma_{\v\r}^{(2)}$ includes the contributions from the interference of one-loop and tree-level scattering 
amplitudes with one extra external particle, and the double real cross-section $d\sigma_{\r\r}^{(2)}$ are 
tree-level contributions with emission of two extra particles. 
The LTD representation of the two-loop scattering amplitude is obtained by setting two 
internal lines on-shell~\cite{Bierenbaum:2010cy}. It leads to the two-loop dual components
$\la \M{0}_N|\M{2}_N(\td{q_i},\td{q_j})\ra$, while the two-loop momenta of the squared one-loop 
amplitude are independent and generate dual contributions of the type 
$\la \M{1}_N(\td{q_i})|\M{1}_N(\td{q_j})\ra$. In both cases, there are two independent
loop three-momenta and $m$ final-state momenta, from where we can reconstruct the 
kinematics of the one-loop corrections entering $d\sigma_{\v\r}^{(2)}$, and the 
tree-level corrections in $d\sigma_{\r\r}^{(2)}$.

\section{NLO corrections to $\gamma^* \to q \bar q (g)$ in four dimensions}
\label{app:4d}

\begin{figure}[htb]
\begin{center}
\includegraphics[width=0.35\textwidth]{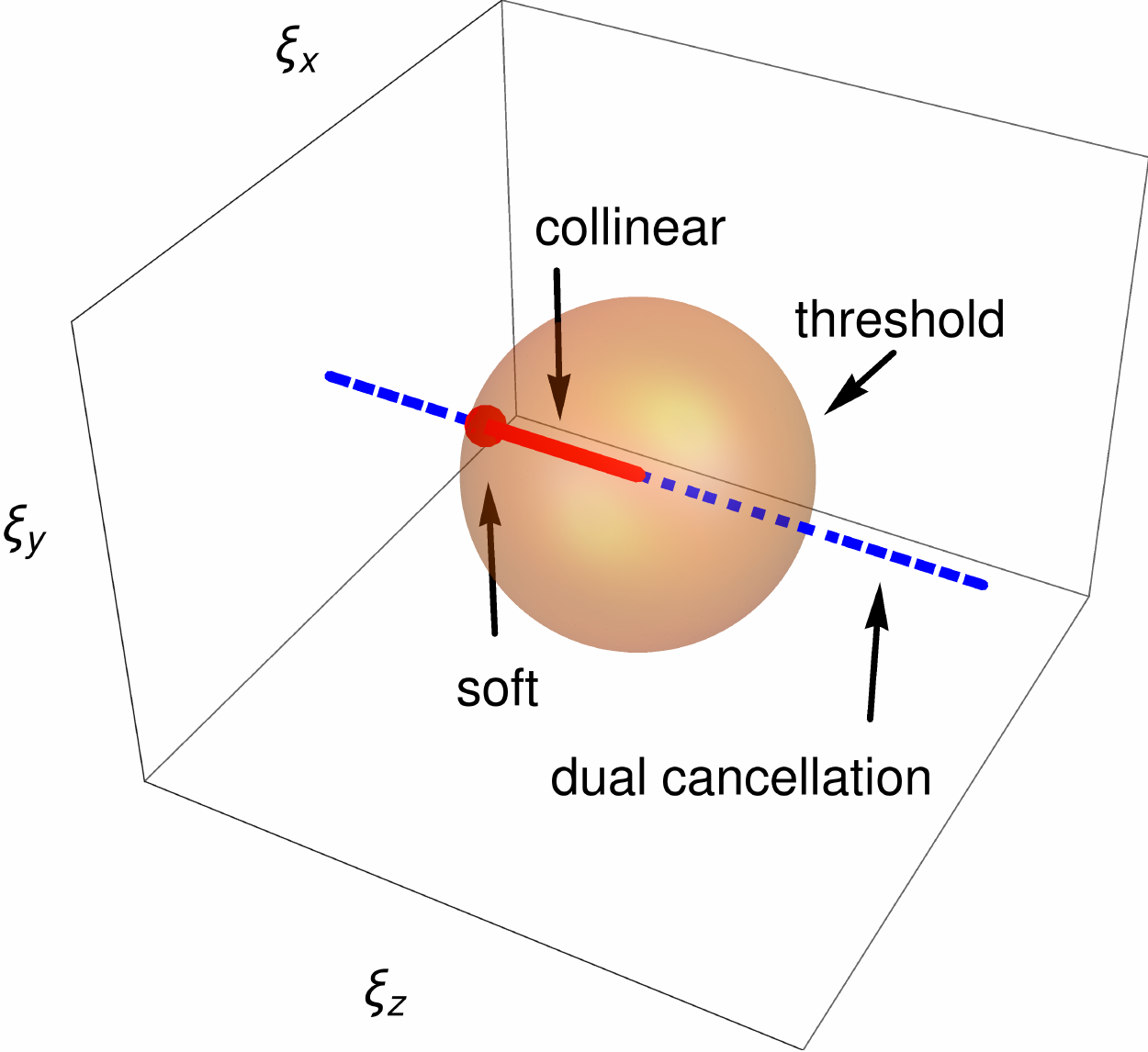} $\qquad$
\includegraphics[width=0.3\textwidth]{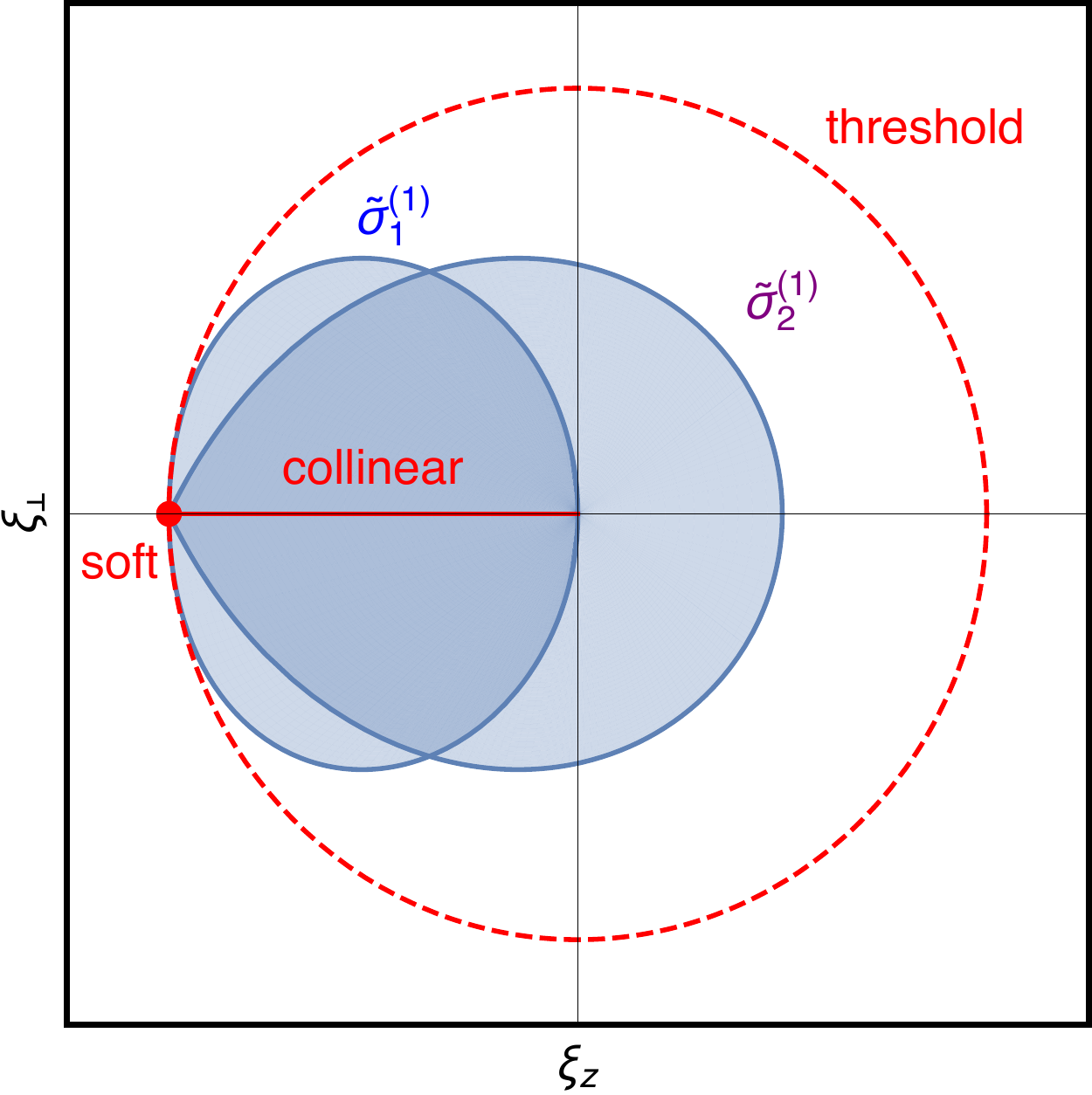}
\caption{Singular structure of the three-point function in the loop coordinates $\ell^{\mu}= \sqrt{s_{12}}/2\, \xi^\mu$ (left plot);
regions in the loop three-momentum space (right plot) where virtual and real kinematics are mapped,
with $\xi_\perp = \sqrt{\xi_x^2+\xi_y^2}$. 
\label{fig:REGIONLightcone}}
\end{center}
\end{figure}

We present in this section the four-dimensional representation of the total cross-section for the
physical process $\gamma^* \to q \bar q (g)$ at NLO. The loop internal momenta are
$q_1=\ell + p_1$, $q_2=\ell+p_{12}$  and $q_3=\ell$, which are parametrised as
\beq
q_i^\mu = \frac{\sqrt{s_{12}}}{2} \, \xi_{1,0} \,  (1, 2\sqrt{v_i(1-v_i)}\,  {\bf e}_{i,\perp}, 1-2v_i)~, 
\eeq
when they are set on-shell, $q_i^2=0$. The direct combination of the virtual and real corrections
leads to the dual cross-section contributions 
\beqn
\nn \widetilde \sigma_{1}^{(1)} &=& \sigma^{(0)} \, \frac{\as}{4\pi} \,  C_F \, \int_0^1 d\xi_{1,0} \, \int_0^{1/2} dv_1 \, 4\, {\cal R}_1(\xi_{1,0},v_1) \, 
\left[ 2 \left( \xiu - (1-v_1)^{-1} \right) - \frac{\xiu (1-\xiu)}{\left(1-(1-v_1) \, \xiu \right)^2} \right]~,  
\label{eq:APPENDIXexplicitsigma1}
\\ \nn \widetilde \sigma_{2}^{(1)} &=& \sigma^{(0)} \, \frac{\as}{4\pi} \,  C_F \, \int_0^1 d\xi_{2,0} \, \int_0^1 dv_2 \, 2 \, {\cal R}_2(\xi_{2,0},v_2) \, 
(1-v_2)^{-1}\,  \Bigg[ \frac{2 \, v_2 \, \xid \left(\xid (1-v_2) - 1\right)}{1-\xid}  \\ &&
- 1 + v_2\, \xid  + \frac{1}{1-v_2 \, \xid} \left(\frac{(1-\xid)^2}{(1-v_2 \, \xid)^2} + \xid^2\right)  \Bigg]~,
\label{eq:APPENDIXexplicitsigma2}
\eeqn
where the integration regions are defined through
\bea
&&{\cal R}_1(\xiu,v_1) = \theta (1-2v_1) \, \theta\left( \frac{1-2v_1}{1-v_1} - \xiu \right)~, \nn \\
&&{\cal R}_2(\xid,v_2) = \theta\left( \frac{1}{1+\sqrt{1-v_2}} - \xid \right)~.
\label{eq:errequeerre}
\eea
A graphical representation of these regions is shown in Fig.~\ref{fig:REGIONLightcone}~(right).
Outside these regions only virtual corrections contribute, and the sum of all the dual integrals 
is finite after the inclusion of suitable UV counter-term. The dual remnant is 
\beqn
\nn \overline \sigma^{(1)}_{\v} &=& \sigma^{(0)} \, \frac{\as}{4\pi} \,  C_F \, \int_0^{\infty} d\xi \, \int_0^1 dv \, 
\Bigg\{ - 2 \, \left(1-{\cal R}_1(\xi,v)\right) \, v^{-1}(1-v)^{-1} \, \frac{\xi^2 (1-2 v)^2+1}{\sqrt{(1+\xi)^2 - 4 v\, \xi}} 
\\ \nn &+& 2 \, \left(1-{\cal R}_2(\xi,v)\right) \, (1-v)^{-1}\,  \left[2 \, v \, \xi \, \left(\xi (1-v) - 1\right)
\left(\frac{1}{1-\xi+\imath 0} + \imath \pi \delta(1-\xi)  \right) - 1 + v \, \xi \right] 
\\ \nn &+& 2\, v^{-1} \left( \frac{\xi (1-v)(\xi (1-2v)-1)}{1+\xi} +1\right) 
- \frac{(1-2v) \, \xi^3 \, ( 12 - 7 m_{\uv}^2 - 4 \xi^2 )} {(\xi^2 + m_{\uv}^2)^{5/2}} \\ 
&-& \frac{2\, \xi^2 (m_{\uv}^2 + 4\xi^2(1-6v(1-v)))}{(\xi^2 + m_{\uv}^2)^{5/2}} \Bigg\}~. 
\label{eq:APPENDIXexplicitremnant}
\eeqn 
In the previous expression, we have identified all the integration variables, 
$\xi=\xid=\xit=\xi_{\uv}$ and $v=v_2=v_3=v_{\uv}$, while $(\xiu,v_1)$ are expressed in terms of 
$(\xid,v_2)$  with the appropriate change of variables~\cite{Sborlini:2016gbr}. 
The last two terms in \Eq{eq:APPENDIXexplicitremnant} that depend on $m_\uv^2$ are the UV counter-terms of the 
vertex and the self-energies.
Putting together the three dual contributions from \Eq{eq:APPENDIXexplicitsigma2} and \Eq{eq:APPENDIXexplicitremnant}, we obtain
\beq
\sigma = \sigma^{(0)} \left( 1 +  3 \, C_F \, \frac{\as}{4\pi}  
+ {\cal O}(\as^2) \right)~,
\label{eq:totalNLO}
\eeq
which agrees with the well-known result available in the literature.
Notice that it was unnecessary to introduce any tensor reduction; Gram determinants 
are naturally avoided in LTD, and therefore the spurious singularities that the tensor reduction 
introduces leading to numerical instabilities in the integration over the phase-space.

\section{Conclusions and outlook}
\label{sec:conclusions}

We have discussed the implementation of a novel algorithm to compute higher-order 
corrections to physical observables. This method is based on the LTD theorem, which states that virtual 
contributions can be expressed as the sum over single-cut (at one-loop) or dual integrals, whose structure closely 
resembles real-emission amplitudes. Moreover, in LTD all the threshold and IR divergences are 
generated in a compact region of the loop three-momentum space. 
The summation over degenerate soft and collinear configurations is achieved
by defining suitable momentum mapping relating the virtual and real kinematics. 
These mappings correlate exactly the kinematical configurations in the integration regions where the IR singularities are originated. 
In the high-energy region unintegrated UV counter-terms cancel the UV singular behaviour, their LTD
representations have been also presented in detail, and admit a natural physical interpretation. 
The algorithm is called unsubtracted because the momentum 
mappings make unnecessary the introduction of IR subtractions. 

These facts can be exploited to perform an integrand-level combination
of real and virtual terms, which leads to a fully local cancellation of singularities 
and allows to implement NLO calculations without making use of DREG.
It constitutes a new paradigm in perturbative calculations because it takes 
advantage of combining directly real and virtual corrections in an integrable four-dimensional 
representation, while providing an easy physical interpretation of the singularities of the scattering amplitudes
and unveil their hidden nature.

The implementation of the method has been illustrated with the simplest physical process 
$\gamma^* \to q \bar{q}(g)$ at NLO in QCD, and explicit expressions in four dimensions have been presented.
We have also succinctly sketched the generalisation to NNLO, and are currently working in the extension 
of the method to deal with massive particles~\cite{massive}. The numerical treatment of threshold 
singularities by contour deformation has also been investigated recently in Refs.~\cite{Buchta:2014dfa,Buchta:2015xda,Buchta:2015wna}

\section*{Acknowledgements}

This work has been supported by CONICET Argentina,
by the Spanish Government and ERDF funds from the European Commission 
(Grants No. FPA2014-53631-C2-1-P and SEV-2014-0398)
and by Generalitat Valenciana under Grant No. PROMETEOII/2013/007. FDM acknowledges suport from Generalitat
Valenciana (GRISOLIA/2015/035). The work of RJHP is partially supported by CONACyT, M\'exico.

\end{document}